\documentclass[12pt]{article}
\usepackage{graphicx}
\usepackage{amsmath}
\usepackage{amssymb}
\usepackage{caption2}
\begin{document}
\bibliographystyle{prsty}
\begin{center}
{\large {\bf \sc{   Structure and decay constant of the $\rho$ meson  with Bethe-Salpeter equation }}} \\[2mm]
Zhi-Gang Wang$^{1}$ \footnote{Corresponding author; E-mail,wangzgyiti@yahoo.com.cn.  } and Shao-Long  Wan$^{2}$   \\
$^{1}$ Department of Physics, North China Electric Power University, Baoding 071003, P. R. China \\
$^{2}$ Department of Modern Physics, University of Science and Technology of China, Hefei 230026, P. R. China \\
\end{center}

\begin{abstract}
In this article, we study  the structure of the  $\rho$ meson
    in the framework of the coupled rainbow Schwinger-Dyson equation and ladder
 Bethe-Salpeter equation with  a confining effective potential.
   The $u$ and $d$ quark propagators get
significantly modified, the mass poles  are
   absent in the timelike region,  which
  implements  confinement naturally.
 The Bethe-Salpeter amplitudes of the $\rho$ meson
  center around  zero momentum and extend to the energy scale about
$q^2=1GeV^2$, which happens to be the energy scale of  chiral
symmetry breaking, strong interactions  in the infrared region
result in bound state. The numerical results of the mass and decay
constant of the  $\rho$ meson are in agreement with the
experimental data.
\end{abstract}

PACS : 14.40.-n, 11.10.Gh, 11.10.St, 12.40.qq

{\bf{Key Words:}}  Schwinger-Dyson equation, Bethe-Salpeter
equation, $\rho$ meson
\section{Introduction}
 Low energy nonperturbative properties of  quantum chromodynamics (QCD)
put forward a great challenge to physicists as the  $SU(3)$ gauge
coupling at low energy scale does not allow  perturbative
calculations. Among the existing theoretical approaches, the coupled
rainbow Schwinger-Dyson equation (SDE) and ladder Bethe-Salpeter
equation (BSE) are typical. They have been very successful in
describing the long distance  properties of  low energy  QCD and the
QCD vacuum (for  reviews, one can consult
Refs.\cite{Roberts94,Tandy97,Roberts00,Roberts03,Alkofer03}). The
SDE can naturally embody  dynamical chiral symmetry breaking and
confinement,  which are two crucial features of QCD, although they
correspond to two very different energy scales
\cite{Alkofer03,Miransky93}. The possible interplay of   two
dynamics lies in the following two facts: the first one is that at
high energy scale the chiral massive quark dresses itself with a
gluon cloud  and quark-antiquark pairs, and creates a constituent
quark mass;  the second one is  the double role of the light
pseudoscalar mesons, as both  Nambu-Goldstone bosons and $q\bar{q}$
bound states. The BSE is a conventional approach in dealing with
two-body relativistic bound state problems \cite{BS51}. From
solutions of the BSE, we can obtain useful information about the
bound state structure of the mesons and obtain powerful tests of the
quark theory.

Many analytical and numerical calculations indicate that the coupled
rainbow SDE and ladder BSE with phenomenological potentials can give
model independent results and
 satisfactory values \cite{Roberts94,Tandy97,Roberts00,Roberts03,Alkofer03}.
The pseudoscalar mesons, especially the $\pi$ and $K$, have been
studied  extensively  to understand the bound state structures  from
 nonperturbative QCD (for example, Ref.\cite{Dai}, for more
literature, one can consult
Refs.\cite{Roberts94,Tandy97,Roberts00,Roberts03,Alkofer03}). Chiral
symmetry and axial Ward identity play an important role, in chiral
limit, the dominating Bethe-Salpeter  amplitude (BSA) of the
pseudoscalar mesons with  the structure $\gamma_5$ is given by
\mbox{$B(q^2)/f_P$},  where $B$ is  the scalar part of the quark
self-energy  and $f_P$ is the decay constant of the pseudoscalar
meson. We can perform many phenomenological analysis without
 solving  the BSE explicitly.

The situation is much worse for the vector mesons. There is a
conserved vector current and the corresponding Ward identity relates
 the longitudinal (not transverse) part of the  vector vertex
with the quark propagator.  We cannot obtain dynamical insight into
the  relevant degrees of freedom without  solving  the BSE directly.
Furthermore, it is more difficult to solve  the vector BSE  than
 the   pseudoscalar BSE due to   a larger  number of
covariants and   higher masses, which require extrapolation  into
larger domain of the complex $q^2$ plane \cite{TandyV}.

The often used effective potential models are confining Dirac
$\delta$ function potential, Gaussian  distribution potential and
flat bottom potential (FBP) \cite{Munczek83,Munczek91,Wangkl93}. The
FBP is a sum of Yukawa potentials, which not only  satisfies gauge
invariance, chiral invariance and fully relativistic covariance, but
also suppresses the singular point which the
 Yukawa potential has. It  displays
  dynamical chiral symmetry breaking and confinement \cite{Wang02},  and works
  well in understanding
  the QCD vacuum (for example, the quark condensate, mixed condensate and vacuum susceptibility) as well as
 the meson  structures (for example, the electromagnetic form factors,
radius and decay constants)\cite{Wangkl93,WYW03,WangWan}.

In this article, we take the point of view that the $\rho$ meson is
  the $q\bar{q}$ bound state, and study  its structure and decay constant
in the framework of the coupled rainbow SDE  and ladder BSE with the
infrared modified FBP, which take  advantages of both the Gaussian
distribution potential and the FBP.

The article is arranged as the followings: we introduce the infrared
modified FBP in section II; in section III, IV and V, we solve the
coupled rainbow SDE and ladder BSE, study  analytical property  of
the quark propagators, finally obtain the mass and decay constant of
the $\rho$ meson; section VI is reserved for conclusion.

\section{Infrared Modified Flat Bottom Potential }
The present techniques in QCD  calculation cannot   give
satisfactory analytical results for the infrared behavior of the
two-point Green's function of the gluon. Infrared enhanced effective
potential models have been phenomenologically quite successful
\cite{Munczek83,Munczek91,Wangkl93} .

One can use a gaussian distribution function to represent the
infrared behavior of the  two-point Green's function of the
gluon\footnote{In this article, we use the metric
$\delta_{\mu\nu}=(1,1,1,1)$, $\left\{\gamma_\mu
\gamma_\nu+\gamma_\nu \gamma_\mu\right\}=2\delta_{\mu\nu}$, the
coordinate $x_\mu=(it,\overrightarrow{x})$, the momentum
$p_\mu=(iE,\overrightarrow{p})$.},
\begin{eqnarray}
  G_{1}(q^2)=\frac{\varpi^2}{\Delta^2}e^{-\frac{q^2}{\Delta}} \,\, ,
\end{eqnarray}
which determines the quark-antiquark interaction through a strength
parameter $\varpi$ and a range parameter $\Delta$ \cite{Munczek91}.
 This form is inspired by the Dirac $\delta$ function potential
(in other words the infrared dominated potential) used in
Ref.\cite{Munczek83}, which it approaches in  limit
$\Delta\rightarrow 0$. The integral $\int d^4q q^{2n}G_{1}(q^2)$ is
finite for  spacelike squared momentum $q^2$ . Such an infrared
behavior can result in large dressing for the quark's
Schwinger-Dyson functions (SDFs) $A(q^2)$ and $B(q^2)$ near $q^2=
0$, the curves at the infrared region may be steep enough to forbid
 extrapolation to   deep timelike region. When we introduce an
extra factor, $q^2/\Delta$, the modified gaussian distribution can
result in more flat curve near   zero momentum. Furthermore,
systematic studies with the coupled SDEs of the quark, gluon and
ghost indicate this type of behavior at about $q^2=0$
\cite{AlkoferSD}. We use the following modified gaussian
distribution to represent the infrared behavior of the two-point
Green's function of the gluon \cite{Munczek91},
\begin{eqnarray}
\frac{\varpi^2}{\Delta^2}e^{-\frac{q^2}{\Delta}}
\rightarrow\frac{\varpi^2}{\Delta^2}\frac{q^2}{\Delta}e^{-\frac{q^2}{\Delta}}
\, \, .
\end{eqnarray}
In numerical calculation, the range parameter  $\Delta$ is taken to
be $\sqrt{\Delta}\approx 0.62GeV$, the Gaussian type of function
$q^2e^{-\frac{q^2}{\Delta}}$ centers around $q=0.6GeV$, and extends
to about $q=1.2GeV$. Systematic studies with the coupled SDEs
indicate that the nonperturbative gluon propagator is greatly
enhanced  at about $q=1GeV$ \cite{AlkoferSD}. The value
$\sqrt{\Delta}\approx 0.62GeV$ is more reasonable than the value
$\sqrt{\Delta}= 0.3GeV$ taken in Ref.\cite{TandyV}.

For   intermediate momentum, we take the FBP as the best
approximation and neglect the contribution  from   perturbative QCD
calculations as the strong  coupling constant at high energy scale
is very small. The FBP is a sum of Yukawa potentials which is an
analogy to   exchange of a series of particles and ghosts with
different masses,
\begin{eqnarray}
G_2(q^{2})&=&\sum_{j=0}^{n}
 \frac{a_{j}}{q^{2}+(N+j \eta)^{2}} \,\, ,
\end{eqnarray}
where $N$ stands for the minimum value of the masses,   $\eta$ is
their mass difference, and   $a_{j}$ is their relative coupling
constant. Definition of momentum regions between infrared and
intermediate momentum is about $\Lambda_{QCD}=200MeV$, which is
 set up naturally by the minimum value of the masses
$N=1\Lambda_{QCD}$. Certainly, there are some overlaps between those
regions, in this way, we can guarantee   continuity  for the
momentum. The FBP  at energy  $N+j\rho$ with $j>3$  extends to the
perturbative region and exhibits  some perturbative  characters. The
infrared modified FBP is supposed to embody  a great deal  of
physical information about  all   momentum regions.

 Due to the particular condition we take for the FBP,
there is no divergence in solving the SDE. In its three dimensional
form, the FBP takes the following form,
\begin{eqnarray}
V(r)&=&-\sum_{j=0}^{n}a_{j}\frac{{\rm e}^{-(N+j \eta)r}}{r} \,\, .
\end{eqnarray}
In order to suppress the singular point at $r=0$, we take the
following conditions,
\begin{eqnarray}
V(0)=constant \, \, , \nonumber \\
\frac{dV(0)}{dr}=\frac{d^{2}V(0)}{dr^{2}}=\cdot \cdot
\cdot=\frac{d^{n}V(0)} {dr^{n}}=0  \,\,  .
\end{eqnarray}
The  $a_{j}$ can be  determined by solving   equations  inferred
from the flat bottom condition. As in  previous literature
\cite{Wangkl93,Wang02,WYW03,WangWan}, $n$ is set to be 9. The
 gluon propagator can be approximated by
\begin{eqnarray}
G(q^{2})&=&G_1(q^{2})+G_2(q^{2}) \, .
\end{eqnarray}

\section{Schwinger-Dyson equation}
The SDE, in effect the functional
 Euler-Lagrange equation of  quantum field theory, provides a natural
  framework for studying the nonperturbative properties  of
  quark and gluon Green's functions. By studying the evolution
  behavior and analytic structure of the dressed quark propagator,
  one can obtain valuable information about  dynamical chiral symmetry
  breaking phenomenon and confinement.
In the rainbow approximation, the SDE takes the following form,
 \begin{equation}
S^{-1}(p)=i\gamma \cdot p + \hat{m}_{u,d}+ 4\pi \int \frac
{d^{4}k}{(2 \pi)^{4}} \gamma_{\mu}\frac{\lambda^a}{2}
S(k)\gamma_{\nu}\frac{\lambda^a}{2}G_{\mu \nu}(k-p)\,\, ,
\end{equation}
where
\begin{eqnarray}
S^{-1}(p)&=& i A(p^2)\gamma \cdot p+B(p^2)\equiv A(p^2)
[i\gamma \cdot p+m(p^2)]\,\, , \\
G_{\mu \nu }(k)&=&(\delta_{\mu
\nu}-\frac{k_{\mu}k_{\nu}}{k^2})G(k^2)\,\, .
\end{eqnarray}

In this article, we assume that a Wick rotation to Euclidean
variables is allowed, and rotate  $p$, $k$ into the Euclidean region
analytically.  Alternatively, one can derive the SDE from
 Euclidean path-integral formulation of the theory, and  avoid
 possible difficulties in performing the Wick
 rotation $\cite{Stainsby}$.  The analytical  structures of quark propagators have
 interesting information about confinement, we will revisit this subject  in  the section V.

\section{Bethe-Salpeter equation}
The BSE is a conventional approach in dealing with  two-body
relativistic bound state problems \cite{BS51}. Precise knowledge
about the quark structure of the $\rho$ meson  can result in better
understanding of its property. In the following, we write down the
ladder BSE of the  $\rho$ meson,
\begin{eqnarray}
 S^{-1}(q+\frac{
P}{2})\chi(q,P)S^{-1}(q-\frac{ P}{2})=-\frac{16 \pi }{3} \int
\frac{d^4 k}{(2\pi)^4}\gamma_\mu \chi(k,P) \gamma_\nu G_{\mu
\nu}(q-k)\,\, ,
\end{eqnarray}
where  $P_\mu$ is  four-momentum of the center of mass of the $\rho$
meson, $q_\mu$ is  relative four-momentum between the two quarks,
 $\gamma_{\mu}$ is the bare quark-gluon vertex,  and
$\chi(q,P)$ is the BSA  of the $\rho$ meson.

We can perform the Wick rotation analytically and continue  $q$, $k$
into  Euclidean region \footnote{To avoid any possible difficulties
in performing the Wick rotation, we can derive  both the SDE and BSE
from   Euclidean path-integral formulation of the theory directly,
then  continue  the four-momentum of the center of mass of the
$\rho$ meson   into   Minkowski
 spacetime analytically, $P^2=-m_\rho^2$. As far as only the numerical values are concerned, the
two approaches are equal.}.
 The Euclidean BSA of the $\rho$ meson can be decomposed as
 \begin{eqnarray}
 \chi(q,P)&=&\epsilon_\mu \chi_\mu(q,P) \, , \\
 \chi_\mu(q,P)&=&\left\{\gamma_\mu - \frac{P_\mu \!\not\!{P}}{P^2}
 \right\} \left\{
 iF_0+\!\not\!{P}F_1-\!\not\!{q}F_2+i[\!\not\!{P},\!\not\!{q}]F_3\right\}
 \nonumber\\
 &&+\left\{q_\mu - \frac{P_\mu q\cdot P}{P^2}
 \right\} \left\{
 F_2+2i \!\not\!{P}F_3 \right\} \nonumber\\
  &&+\left\{q_\mu - \frac{P_\mu q\cdot P}{P^2}
 \right\} \left\{F_4+i\!\not\!{P}F_5-i\!\not\!{q}F_6+[\!\not\!{P},\!\not\!{q}]F_7 \right\}
 \, ,
 \end{eqnarray}
 due to  Lorentz covariance. Here $\epsilon_\mu$ is the
 polarization vector of the $\rho$ meson \cite{VectorBS}.
The BSAs $F_{i}(q,q\cdot P,P)$ can be expanded  in terms of
Tchebychev polynomials $T^{\frac{1}{2}}_{n}(\cos \theta)$,
\begin{eqnarray}
F_{i}(q,q\cdot P,P)=\sum_0^{\infty}i^nF_{i}^{n}(q,P) q^n P^n
T^{\frac{1}{2}}_{n}(\cos \theta)\,\, ,
\end{eqnarray}
where  $\theta$ is  the included  angle between  $q_\mu$ and
$P_\mu$. It is impossible to solve an infinite series of coupled
equations of the $F_{i}^n$, we have to make truncation in one or the
other ways. Numerical calculations indicate that taking only some
terms with $n=0,1,2$  can give satisfactory results \footnote{We can
borrow some ideas from the twist-2 light-cone distribution
amplitudes $\phi(\mu,u)$ of the pseudoscalar mesons $\pi$ and $K$,
where  $u$ is  momentum fraction of the quark, and $\mu$ is the
energy scale. The $\phi(\mu,u)$ is always expanded in terms of
Gegenbauer polynomials, $\phi(\mu,u)=6u(1-u)\left\{
1+a_1(\mu)C_1^{3/2}(2u-1)+a_2(\mu)C_2^{3/2}(2u-1)+a_4(\mu)C_4^{3/2}(2u-1)+\cdots\right\}$,
where   $ C_1^{3/2}(2u-1)$, $ C_2^{3/2}(2u-1)$
 and $ C_4^{3/2}(2u-1)$ are  Gegenbauer polynomials, and
  $a_i(\mu)$ are  nonperturbative coefficients.
 The coefficients $a_i(\mu)$ can be estimated with the
 QCD sum rules approach,  for large $i$,  the $a_i(\mu)$ involve
 high dimension vacuum condensates which are known poorly. In
 general, one can retain only the first few terms and fit them with
 experimental data, and the truncated light-cone distribution
amplitude $\phi(\mu,u)$ always gives satisfactory results
\cite{LCSRA}.
 In this article, we retain
 only some  terms with $n=0$, $1$ and $2$
 in   expansion with   Tchebychev polynomials $T^{1/2}_{n}(\cos
\theta)$, the contributions from other terms are supposed to be
small and neglected here. If the contributions from the  neglected
terms are large, they cannot decouple approximately from the BSEs
and warrant  the BSEs have reasonable solution. The numerical
results indicate that the dominating contribution comes from the
$F_0^0$, the sub-dominating contributions come from the $F_1^0$ and
$F_4^0$, and $F_0^0 \gg q^2 m_\rho^2 F_0^2$, we expect the
truncation is reasonable.
     It is obvious that if we take into
account more terms in expansion, more accurate values can be
obtained. However, the analytical expressions of the coupled BSEs
become very clumsy and are beyond  capability of our computer in
numerical calculations. },
\begin{eqnarray}
 \chi_\mu(q,P)&=&\left\{\gamma_\mu - \frac{P_\mu \!\not\!{P}}{P^2}
 \right\} \left\{
 iF_0^0+i[4(q \cdot P)^2-q^2P^2]F_0^2+\!\not\!{P}F_1^0-\!\not\!{q}q\cdot P F_2^1+i[\!\not\!{P},\!\not\!{q}]F_3^0\right\}
 \nonumber\\
 &&+\left\{q_\mu - \frac{P_\mu q\cdot P}{P^2}
 \right\} \left\{
 q\cdot P F_2^1+2i \!\not\!{P}F_3^0 \right\} \nonumber\\
  &&+\left\{q_\mu - \frac{P_\mu q\cdot P}{P^2}
 \right\} \left\{F_4^0+i\!\not\!{P} q \cdot P F_5^1-i\!\not\!{q}F_6^0+[\!\not\!{P},\!\not\!{q}]F_7^0
 \right\} \,.
 \end{eqnarray}

In solving the BSE, it is important to translate the BSAs
$F_{i}^{n}$ into the same  dimension of mass to facilitate the
calculation,
\begin{eqnarray}
&&F_{0}^{0}\rightarrow \Lambda^{0}F_{0}^{0}, \, F_{0}^{2}\rightarrow
\Lambda^{2}F_{0}^{2}, \, F_{1}^{0}\rightarrow \Lambda^{1}F_{1}^{0},
\,F_{2}^{1}\rightarrow \Lambda^{3}F_{2}^{1}, \nonumber\\
 &&F_{3}^{0}\rightarrow \Lambda^{2}F_{3}^{0} \, , F_{4}^{0}\rightarrow \Lambda^{1}F_{4}^{0}, \,F_{5}^{1}\rightarrow
\Lambda^{4}F_{5}^{1}, \, F_{6}^{0}\rightarrow \Lambda^{2}F_{6}^{0},
\, \nonumber \\
 &&F_{7}^{0}\rightarrow \Lambda^{3}F_{7}^{0}, \, q\rightarrow
q/\Lambda, \, P\rightarrow P/\Lambda \, , \nonumber
\end{eqnarray}
where  $\Lambda$ is a quantity with dimension of mass.
 The ladder BSE of the $\rho$ meson
 can be projected into the following nine coupled integral equations,
\begin{eqnarray}
\sum_j H(i,j)F_j^{0,1,2}(q,P)&=&\sum_j \int d^4k K(i,j) \, ,
\end{eqnarray}
where $H(i,j)$ and $K(i,j)$ are $9\times 9 $  matrices, the
corresponding ones for the pseudoscalar mesons are $4\times 4 $
matrices \cite{WYW03}. The analytical expressions of the matrix
elements $H(i,j)$ and $K(i,j)$ are cumbersome and will take up more
than seven pages, and not shown explicitly for simplicity.

We can introduce a parameter $\lambda(P^2)$ and solve  above
equations as an eigenvalue problem.  If there really exist a bound
state in the vector channel, the mass of the $\rho$ meson can be
determined by the condition $\lambda(P^2=-m_{\rho}^2)=1$,
\begin{eqnarray}
\sum_j H(i,j)F_j^{0,1,2}(q,P)&=&\lambda(P^2)\sum_j \int d^4k K(i,j)
\, \, .
\end{eqnarray}

The matrix elements $H(i,j)$  are functions of the quark's SDFs
$A(q^2+\frac{P^2}{4}\pm q \cdot P)$ and $B(q^2+\frac{P^2}{4}\pm q
\cdot P)$ . The relative four-momentum $q_\mu$ is a quantity in
Euclidean spacetime,  while the center of mass four-momentum $P_\mu$
is a quantity in   Minkowski spacetime,  $q \cdot P$ varies
throughout a complex domain.  We can expand the $A$ and $B$ in terms
of Taylor series of  $q \cdot P$ to avoid solving the SDE with
complex values of  quark momentum, for example,
\begin{eqnarray}
A(q^2+\frac{P^2}{4}\pm q \cdot P)&=&A(q^2+\frac{P^2}{4})\pm
A(q^2+\frac{P^2}{4})' q \cdot P+\cdots \,\, . \nonumber
 \end{eqnarray}
The other problem is that we cannot solve the SDE in the timelike
region.  The two-point   Green's function of the gluon cannot be
exactly inferred from the $SU(3)$ gauge theory even in  small
spacelike momentum region. We can extrapolate the values of the
$A(q^2)$ and $B(q^2)$ from the spacelike region smoothly to the
timelike region $q^2=-\frac{m_\rho^2}{4}\approx-0.15GeV^2$ with
suitable polynomial. The masses of the vector mesons are larger than
the pseudoscalar mesons. So it is very difficult to extrapolate the
values to  the  deep timelike region. We must be careful in choosing
the polynomial to avoid possible violation of confinement in sense
of   appearance of pole masses $q^2=-m^2(q^2)$ in the timelike
region \cite{Munczek91,WYW03}. This requires a certain amount of
fine tuning. Furthermore, if the curves of the $A(q^2)$ and $B(q^2)$
are very steep near $q^2=0$, very large values of the
$A(q^2+\frac{P^2}{4})$ and $B(q^2+\frac{P^2}{4})$ are obtained in
the timelike region ($q^2+\frac{P^2}{4}\prec 0$), the solution of
the BSE in the small spacelike region ($0\preceq q^2\prec
\frac{m_\rho^2}{4}$) is not reasonable. In this article, we use the
modified Gaussian distribution rather than the Gaussian distribution
to modify  the infrared behavior of
 the two-point green's function of the gluon to outcome  the above
 difficulties.

 Finally we write down   normalization condition for
the BSAs of the $\rho$ meson,
\begin{eqnarray}
 \frac{N_c}{3}\int \frac{d^4q}{(2\pi)^4} Tr \left\{ \bar{\chi}
\frac{\partial S^{-1}_{+}} {\partial P_{\mu}}\chi(q,P) S^{-1}_{-}
+\bar{\chi} S^{-1}_{+} \chi(q,P) \frac{\partial S^{-1}_{-}}
{\partial P_{\mu}} \right\}=2 P_{\mu}\, \, ,
\end{eqnarray}
where $\bar{\chi}=\gamma_4 \chi^+ \gamma_4$, $S_+=S(q+\frac{ P}{2})$
and $S_-=S(q-\frac{ P}{2})$.

\section{Coupled rainbow SDE and ladder BSE}
Now we study  the coupled equations of the rainbow SDE and ladder
BSE of the $\rho$ meson.

In order to demonstrate   confinement of quarks, we  take the
Fourier transform with respect to   Euclidean time $T$
 for the  scalar part  of the quark propagator \cite{Roberts94,Roberts00,Maris95},
 \begin{eqnarray}
 S^{*}_{s}(T) & =&  \int_{-\infty}^{+ \infty} \frac{dq_{4}}{2 \pi} e^{iq_{4}T}
 \frac{B(q^2)}{q^2A^2(q^2)+B^{2}(q^2)}|_{ \overrightarrow{q}=0}\,\, ,
 \end{eqnarray}
where the 3-vector part of $q_\mu$ is set to zero.
 If  the $S(q)$ has a mass pole at $q^2=-m^2(q^2)$ in the  real timelike region, the Fourier transformed
  $S^{*}_{s}(T)$ would fall off as $e^{-mT}$ for large $T$.
In   numerical calculations, for small $T$, the values of
$S^{*}_{s}$ are positive,  and  decrease rapidly to zero  with
increase  of  $T$, which are compatible with the result (tendency of
curve  with respect to $T$) from  lattice simulations
\cite{Bhagwat03}. For large $T$, the values of $S^{*}_{s}$ are
negative, except for occasionally a very small fraction of positive
values.  The negative values of $S^{*}_{s}$ indicate  an explicit
violation of   axiom of reflection positivity \cite{Jaffee},
   the quarks are not physical observable.

The $u$ and $d$ quarks have small current masses, the dressing or
renormalization is large and the curves of the SDFs are  steep,
which indicates  dynamical chiral symmetry breaking phenomenon.
 At zero momentum, $m_u(0)=m_d(0)=0.51 GeV $, the Euclidean constituent quark masses are
   $m_u(m_u)=m_d(m_d)=0.42 GeV $,  which
  are compatible with the constituent quark masses
in  literature. From the solutions of BSEs of  the $\rho$  meson as
an eigenvalue problem, we  obtain the mass,
\begin{eqnarray}
m_{\rho}&=&770MeV \,\, .
\end{eqnarray}
It is obvious
\begin{eqnarray}
m_u(m_u)+m_d(m_d) >m_\rho \, .
\end{eqnarray}
The attractive interaction between the quark and antiquark in the
infrared region can result in   bound state.

\begin{figure}
 \centering
 \includegraphics[totalheight=7cm,width=7cm]{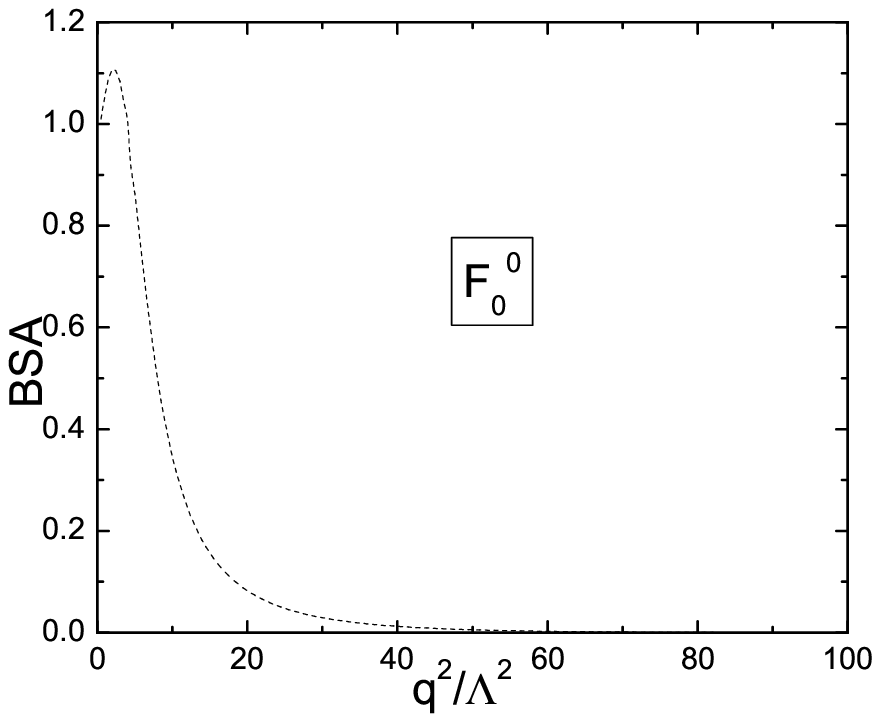}
 \includegraphics[totalheight=7cm,width=7cm]{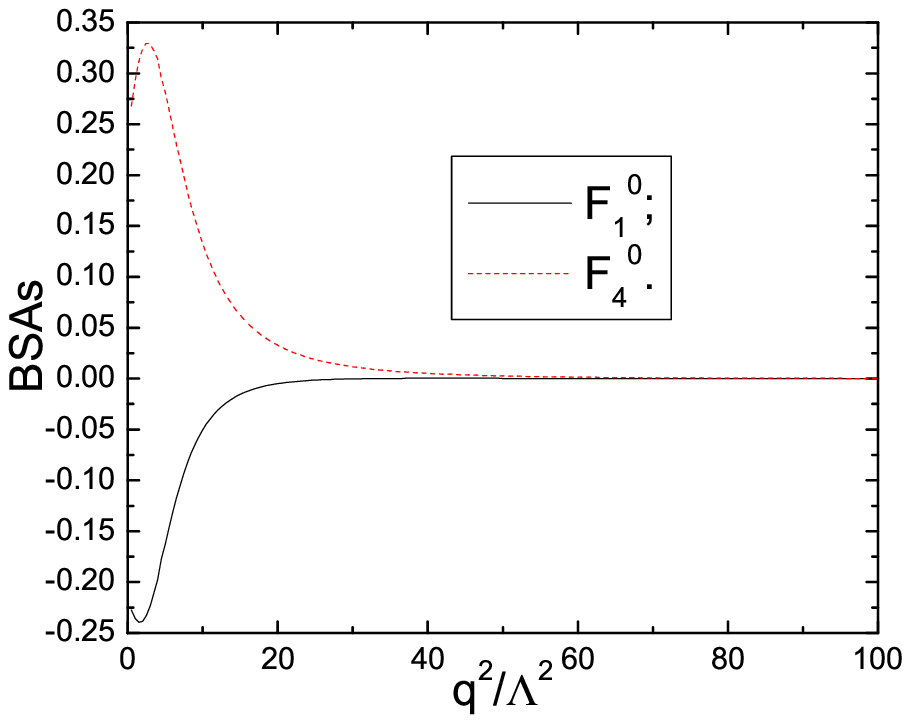}
 \caption{The   dominating and sub-dominating components of the BSAs . }
\end{figure}

\begin{figure}
 \centering
 \includegraphics[totalheight=7cm,width=8cm]{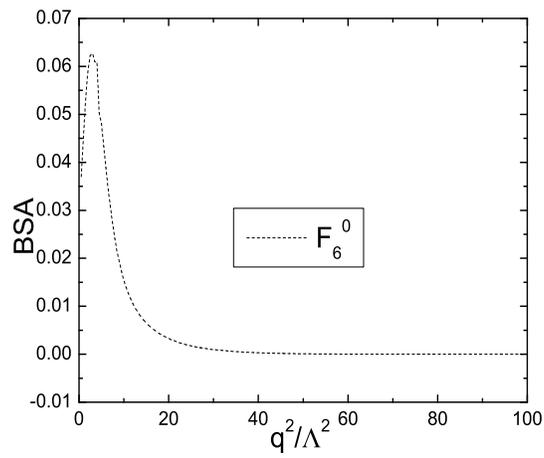}
  \caption{One of the small components of the BSAs  . }
\end{figure}
 The dominating and sub-dominating components of the BSAs are the $F_0^0$, $F_1^0$ and
 $F_4^0$, the six components $F_0^2$, $F_2^1$,  $F_3^0$,
$F_5^1$, $F_6^0$ and $F_7^0$   are very small and of minor
importance.
 From  plotted BSAs (see Fig.1 and Fig.2 as examples), we can see
that  the BSAs of the $\rho$ meson  have  analogical momentum
dependence, while
   the quantitative values are different from each other. Just like the $\bar{q}q$ ,
   $\bar{q}Q$ and $\bar{Q}Q$
   pseudoscalar  mesons \cite{WYW03},
   the  BSAs of the $\rho$ meson
center around  zero momentum and  extend to the energy scale about
$q^2=1GeV^2$, which happens to be the energy scale of  chiral
symmetry breaking. The strong interactions in the infrared region
result  in bound state. The BSAs of the $\rho$  meson  can give
satisfactory  value for the decay constant,  which is defined by
\begin{eqnarray}
 f_{\rho}m_\rho \epsilon_\mu &=& \langle0|\bar{q}\gamma_\mu   q |\rho(P)\rangle \, , \nonumber \\
&=& N_c \int Tr \left[\gamma_\mu  \chi(k,P)\right] \frac{d^4
k}{(2\pi)^4} \, .
\end{eqnarray}
Carrying  out trace explicitly, we can see that only the BSAs
$F_0^0$ and $F_6^0$ are relevant to the decay constant. The $F_6^0$
is numerically very small  (see Fig.2), the dominating contribution
comes from the $F_0^0$. Finally we obtain the value of the decay
constant,
\begin{eqnarray}
f_{\rho}=223 MeV    \,\, ,
\end{eqnarray}
which is in agreement with the  experimental data.

In calculation, the input parameters are taken as $N=1.0 \Lambda $,
$V(0)=-14.0 \Lambda$,
 $\eta=5.0\Lambda$, $\hat{m}_{u}=\hat{m}_{d}=6 MeV$,  $\Lambda=200 MeV$, $\varpi=1.3 GeV$ and  $\Delta=0.39 GeV^2$.

\section{Conclusion }
In this article, we study  structure of the $\rho$ meson in the
framework of the coupled rainbow SDE and ladder BSE with the
confining effective potential. By  solving  the coupled rainbow SDE
and ladder BSE as an eigenvalue problem numerically, we obtain the
SDFs, BSAs, mass and decay constant of the $\rho$ meson.

The dressing (or renormalization) for the SDFs of the $u$ and $d$
quarks is large  and the curves are steep, which indicate  dynamical
chiral symmetry breaking phenomenon. The mass poles  are absent in
the timelike region, which implements   confinement naturally. The
BSAs of the $\rho$ meson  have  analogical momentum dependence,
while
   the quantitative values are different from each other, where the  dominating and sub-dominating
components are the $F_0^0$, $F_1^0$ and $F_4^0$, other six
components $F_0^2$, $F_2^1$, $F_3^0$, $F_5^1$, $F_6^0$, $F_7^0$ are
of minor importance. The  BSAs  center around zero momentum and
extend to the energy scale about $q^2=1GeV^2$, which happens  to be
the energy scale of
  chiral symmetry breaking.  Strong interactions in the
infrared region result  in  bound  state.

The numerical results of the mass and decay constant of the $\rho$
meson are in agreement with the  experimental data. Once
satisfactory SDFs and BSAs of  the $\rho$ meson are obtained, we can
use them in other phenomenological analysis.

\section*{Acknowledgment}
This  work is supported by National Natural Science Foundation,
Grant Number 10405009,  and Key Program Foundation of NCEPU. The
main work comes from  fruitful discussions with Dr. H. Q. Zhou at
IHEP.

\end{document}